\documentclass[preprint,10pt,twocolumn,5p]{article}

\usepackage{graphicx}

\usepackage{amssymb}

\usepackage{xspace}
\usepackage{amsmath}
\usepackage{hyperref}
\newcommand{\hs}{\ensuremath{h_s}\xspace}
\newcommand{\n}{\ensuremath{\eta}\xspace}
\newcommand{\V}{\ensuremath{V}\xspace}
\newcommand{\gamman}{\ensuremath{\gamma_n}\xspace}
\newcommand{\us}{\ensuremath{\mu_s}\xspace}
\newcommand{\uk}{\ensuremath{\mu_k}\xspace}

\newcommand{\hij}{\ensuremath{h_{ij}}}

\newcommand{\nij}{\ensuremath{\hat{n}_{ij}}}
\newcommand{\tij}{\ensuremath{\hat{t}_{ij}}}

\newcommand{\ri}{\ensuremath{\vec{r}_{i}}}
\newcommand{\rj}{\ensuremath{\vec{r}_{j}}}
\newcommand{\vi}{\ensuremath{\vec{v}_{i}}}
\newcommand{\vj}{\ensuremath{\vec{v}_{j}}}
\newcommand{\vij}{\ensuremath{\vec{v}_{ij}}}

\newcommand{\vpi}{\ensuremath{\vec{v}_{pi}}}
\newcommand{\vpj}{\ensuremath{\vec{v}_{pj}}}
\newcommand{\wi}{\ensuremath{\vec{\omega}_{i}}}
\newcommand{\wj}{\ensuremath{\vec{\omega}_{j}}}
\newcommand{\vpij}{\ensuremath{\vec{v}_{pij}}}
\newcommand{\vpji}{\ensuremath{\vec{v_p}_{ji}}}
\newcommand{\tr}{\ensuremath{\theta_r}}
  
\newcommand{\dt}{\ensuremath{\delta t}\xspace}

\newcommand{\Q}{\ensuremath{\mathbf{q}}}
\newcommand{\qcero}{\ensuremath{q_0}}
\newcommand{\quno}{\ensuremath{q_1}}
\newcommand{\qdos}{\ensuremath{q_2}}
\newcommand{\qtres}{\ensuremath{q_3}}

\begin{document}
\bibliographystyle{cpc}



\title{Oedometric Test, Bauer's Law and the Micro-Macro Connection for
  a Dry Sand}

\author{
  W.~F.~Oquendo$^{1,3}$\thanks{Corresponding author: wfoquendop@unal.edu.co }, J.~D.~Mu\~noz$^{1,3}$ and A.~Lizcano$^{2,3}$ \\
  \small $^1$ Simulation of Physical Systems Group, Department of Physics,\\
  \small  Universidad Nacional de Colombia, Carrera 30 No. 45-03,
  \small  Ed. 404, Of. 348, Bogota DC, Colombia\\
  \small $^2$ Department of Civil and Enviromental Engineering, \\
  \small  Universidad de los Andes, Carrera 1 No. 18A-10, Bogota DC, Colombia\\
  \small $^3$ Centre for Basic and Applied Interdisciplinary Studies in
  Complexity\\
  \small CEiBA-Complejidad, Carrera 1 No. 18A-10, Bogota DC, Colombia
}

\maketitle



\begin{abstract}
  What is the relationship between the macroscopic parameters of the
  constitutive equation for a granular soil and the microscopic forces
  between grains?  In order to investigate this connection, we have
  simulated by molecular dynamics the oedometric compression of a
  granular soil (a dry and bad-graded sand) and computed the
  hypoplastic parameters $h_s$ (the granular skeleton hardness) and
  $\eta$ (the exponent in the compression law) by following the same
  procedure than in experiments, that is by fitting the Bauer's law
  $e/e_0 = \exp(-(3p/h_s)^n)$, where $p$ is the pressure and $e_0$ and
  $e$ are the initial and present void ratios. The micro-mechanical
  simulation includes elastic and dissipative normal forces plus slip,
  rolling and static friction between grains. By this way we have
  explored how the macroscopic parameters change by modifying the
  grains stiffness, $V$ ; the dissipation coefficient, $\gamma_n$; the
  static friction coefficient, $\mu_s$; and the dynamic friction
  coefficient, $\mu_k$ . Cumulating all simulations, we obtained an
  unexpected result: the two macroscopic parameters seems to be
  related by a power law, $h_s =0.068(4)\eta^{9.88(3)}$. Moreover, the
  experimental result for a Guamo sand with the same granulometry fits
  perfectly into this power law. Is this relation real? What is the
  final ground of the Bauer's Law? We conclude by exploring some
  hypothesis.

\end{abstract}





\section{Introduction}
Granular media are present everywhere. Examples go from common salt at
the kitchen to planetary ice rings. In tons amount, granular media are
the second most used materials, after water~\cite{gennespowderbook,
degennes}. One of the most interesting examples of granular media are
soils. A good understanding of the behavior of soils under several
conditions is determinant in terms of engineering design, building
planning and construction processes. Furthermore, soils and granular
media represents a new paradigm in physics, and simulations by
computers have turned out to be an excellent tool to gain deep insight
in their behavior.

Traditionally, two main streams have been used to understand soils
\cite{powders2005, hipo1, hypo06}. On one hand, engineers propose
macroscopic constitutive relations in order to reproduce the
deformation (or deformation rate) in terms of the strain (or strain
rate) of the soil. Many formulations, like viscoplasticity, plasticity
and hypoplasticity have been successful to reproduce the experimental
results. For instance, the hypoplastic model has been very useful to
reproduce the experimental behavior of dry sands under monotonic
loads. This model uses eight parameters to characterize the soil, and
all of them can be obtained from element tests. On the other hand,
physicists try to understand the soil behavior as the global result of
microscopic forces between grains~\cite{powders2005,
  evesque01, evesque02, ratcheting1}. This global behavior is usually
investigated by means of computer
simulations~\cite{rapaportGranChapter}.  To determine the relationship
between the macroscopic parameters of the constitutive equation for a
granular soil and the microscopic forces between grains is one of the
main questions in the field.

In this work we explore this connection for the case of a low
polydisperse (bad-graded) dry sand when modeled by the hypoplastic
theory. In particular, we want to investigate the dependence of two
hypoplastic parameters that are obtained from the oedometric test on
the soil: the granular skeleton hardness, \hs, and the exponent in the
compression law, \n, in terms of the parameters governing the
microscopic interactions between grains. For this purpose we perform
three-dimensional discrete element simulations of this element test on
a dry sand of spherical grains for several combinations of four
microscopic parameters, namely: the stiffness of the grains \V, the
normal damping coefficient \gamman, the kinetic friction coefficient
\uk and the static friction coefficient \us. The software, developed
by us, also includes rolling forces and torques, and therefore it is
able to reproduce global reorganizations by rolling. The microscopic
parameters are varied around those for a Guamo sand \cite{Guamo}. A
simulated oedometric test is performed for each set of microscopic
parameters, and the two hypoplastic parameters are measured from the
simulation for each case.  Finally, all simulations are put together
in order to intend (if any) an empirical relation between the two
macroscopic parameters, and the results are compared with the
experimental values of the hypoplastic parameters for a Guamo sand.
Section \ref{microforces} introduces the microscopical forces
included. Section \ref{IntMethods} shows the integration algorithms
employed. The simulated oedometric test are performed and analyzed in
Sec. \ref{SimOedometric}.  Finally, Section \ref{Conclusions}
summarizes the conclusions and introduces suggestions for further
research.

\section{The microscopic forces and torques}\label{sec:micforcemodel}
\label{microforces}
\begin{figure}[ht]
  \centering
  \includegraphics[scale=0.45]{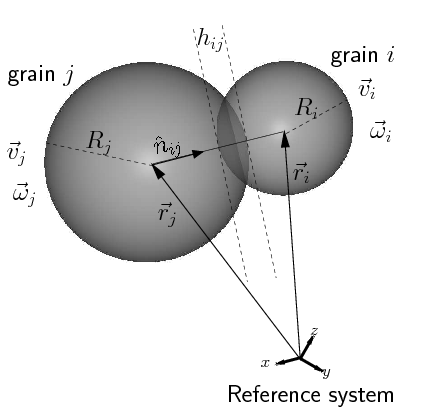}
  \caption{Three-dimensional vector quantities for two spherical
    grains in contact.}\label{fig:balls}  
\end{figure}

The experimental system we want to simulate is a dry sand with very
low polydispersity (between 0.85 and 1.15mm in diameter). 
We model the grains  as spherical
particles in three dimensions. The torques and forces between grains
act in normal and tangential directions and dissipate energy on both
of them. In the
following, the subindexes $i$ and $j$ represent particle $i$ and
$j$, respectively, and the subindex ${ij}$ denotes relative
quantities. 
(see Figure~\ref{fig:balls} and table~\ref{tbl:forceParam} for
details).

\begin{table}
  \centering
  \begin{tabular}{cc}\hline\hline
    \textbf{Parameter} & \textbf{Definition}\\\hline\hline
    $R_{i(j)}$ & Radius of particle $i(j)$\\
    $z_{i(j)}$ & Position of the contact point $i(j)$\\
    $R_{i(j)}' = R_{i(j)} - z_{i(j)} $ & Corrected Radius of particle
    $i(j)$\\
    $V_i = \dfrac{E_i}{1 - \nu^2}$ & Stiffness of grain $i$ \\
    $E_i$ & Young modulus of grain $i$\\
    $\nu_i$ & Poisson modulus of grain $i$\\
    $\nij$ & Normal unitary vector\\
    $\tij$ & Normal tangential vector\\
    $\Vec B^n$ & Normal component of vector $\Vec B$\\
    $\Vec B^t$ & Tangential component of vector $\Vec B$\\
    $m_{ij}$ & Reduced mass\\
    $\theta_r$ & Rolling angle\\\hline 
  \end{tabular}
  \caption{Parameter definitions of the microscopic
    model}\label{tbl:forceParam} 
\end{table}

The total force in normal direction is given by
\begin{align}
  \Vec F_{ij}^n = 
  \frac{4}{3} 
  \sqrt{\frac{R_i R_j}{R_i + R_j}}
  \frac{V_i V_j}{V_i + V_j} & h_{ij}^{3/2}\nij\notag\\
  &-  \frac{m_im_j}{m_i+
  m_j}\gamma_n\sqrt{\hij}\vij^n,
\end{align}
where the first term is the Hertz elastic force and the second one is
a normal damping force that reproduces the experimental effect of the
restitution coefficient \cite{restibrilliantov,
  reviewnormalforces}. The variable $\hij = R_i + R_j - |\vec r_i -
\vec r_j|$ is the inter-penetration distance between particles. The
normal unitary vector is computed as\ \ $\nij = (\ri - \rj)/|\ri -
\rj|$. In all cases but static friction, the tangential unitary vector
is computed as $\tij = \vij^t/|\vij^t|$, where $\vij^t$ is the
tangential component of the relative velocity.  In addition, in the
normal direction we have a torque that slows down the relative angular
rotations on the normal direction. As an alternative to the
traditional Cundall-Strack approach~\cite{herrmannluding,ludingobjec},
we derived a new expression for this torque 
\cite{oquendoThesesMsc}. It reads
\begin{equation}\label{equ:ttor}
  \vec\tau_{ij} = -4\pi\gamma_{nt}%
  m_{ij}R_{ij}\hij\vec\omega_{ij}^n.
\end{equation}
This expression can be deduced in the following way: when two grains
are touching each other and have a non-null relative angular velocity
on the normal direction, the contact surface of one grain rotates
relative to the other one. For each area element on the contact
surface we assign a kinetic friction force that opposes to the
relative motion. The net sum of this forces is zero, since they
cancels in pairs, but the torque does not cancel. By summing up all
torque contributions on the contact surface, we obtain
\eqref{equ:ttor}. This expression  depends on the reduced
radius and the penetration depth as a consequence of the geometry of
the contact surface without additional assumptions.

On the tangential direction, the forces acting between grains depend
on the relative position and motion. Furthermore, in order to compute
the torque we need to define the location of the contact point
\textit{c}, i.e. the point where the forces are applied. It is usual
to define \textit{c} to be in the middle of the contact surface,
despite the polydispersity of the sample, but one can show by simple
geometrical arguments that this point \textit{c} is located
at~\cite{oquendoThesesMsc}
\begin{equation} 
  z_{i(j)} = \hij \frac{R_{j(i)}}{R_i + R_j}.
\end{equation} 
The value $ z_{i(j)}$$=$$\hij/2$ requires $R_i$$=$$R_j$.  

In order to compute the sliding friction force and torque, we check
out 
the value of the tangential relative sliding velocity at the contact
point,
\begin{equation}
  \vpij = \vpi - \vpj = \vi - \vj - (R_i'\wi + R_j'\wj)\times
  \nij. 
\end{equation}
If this velocity is different from zero, the particles are
sliding and we apply the following force and
torque: 
\begin{align} 
  \Vec F^t_{ij}  = -\mu_k |\Vec F^n| \tij ,\\
  \vec\tau_{ij} = 
  -R_i'\nij\times\Vec F^t_{ij}.
\end{align} 
If the sliding velocity is almost null (in fact, less than a
  small predefined velocity) we compute the objective
rolling velocity~\cite{ludingobjec} as 
\begin{equation}\label{equ:vptrolling}
  \vpij = \vpji = \vpij^t = R_{ij}'(\vec\omega_i - \vec\omega_j)\times\nij,
\end{equation} 
where $R_{ij}'$ is the reduced effective radius. If this velocity is
non-null, the particles are rolling. By extending 
the rolling model to three dimensions and applying on two
soft spherical bodies we obtain the following expression for
the rolling friction force and torque~\cite{oquendoThesesMsc}, 
\begin{align}
\Vec F^t_{ij} & = - 
  \frac{|\Vec F^n|\tan\tr}{1 + I/m_i{R_i'}^2} \ \tij,\\
  \vec\tau_{ij} & = \frac{R_{i}'|\Vec F^n|\tan\tr}{1 +
m_{i}{R_{i}'}^2/I} \ \tij\times\nij,
\end{align}
where the rolling constraint is clear.
Finally, if the particles are not sliding and not rolling over each
other, they are in 
relative rest. But, if there is a tangential relative force, the
particles could not stay in rest unless a tangential static friction
force is present. In this case the tangential unitary vector $\tij$ is 
computed from the tangential force as \mbox{$\tij = \Vec F^t / |\Vec
F^t|$}. For the static friction force we apply a simplified
model~\cite{frictionmodels01},
\begin{equation}\label{equ:staticfriction}
  \Vec F_{ij}^t = 
  -\Vec F^t,  \text{if $|\Vec F^t| < \mu_s |\Vec F_{ij}^n|$}. 
\end{equation} 

With the model presented so far we were able to reproduce complex
behaviors as sliding or rolling and dissipation on both normal and
tangential directions. We tested each force implementation by making
particular simulations for each of the following cases: sliding,
rolling, only static friction and normal
dissipation. For each case the model worked properly and also for the
case with all the interactions turned on.

\section{Integration methods}
\label{IntMethods}
The simulation method we used is Molecular Dynamics (MD), also known
as Discrete Element Method (DEM). In MD, the time evolution is traced
on discrete time steps of size \dt. The size of \dt depends on the
mechanical and geometrical properties of the system and is typically
of the order $O(\dt) \simeq 10^{-5} - 10^{-7}$ s. The MD computes the
next positions and velocities (or the next angular orientation and
angular velocities) by solving the second Newton law for each particle
in terms of the current positions, velocities and forces (or the
current orientation, angular velocities and torques). This needs not
only a model for the forces and torques, as given in
Section~\ref{sec:micforcemodel}, but also an integration algorithm.

\begin{figure}[ht]
  \centering
  \includegraphics[scale=0.40]{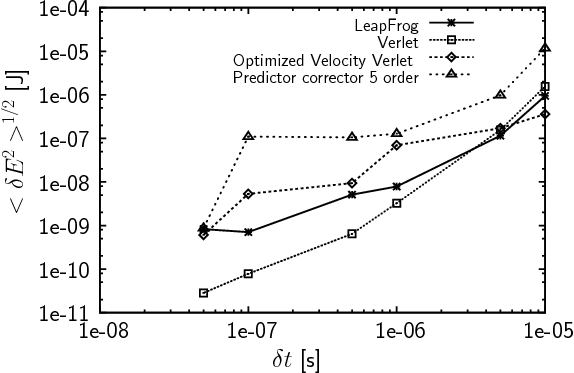}
  \caption{Variation of the total mechanic energy as a function of the
  time step \dt for different translation integration
  algorithms.}\label{fig:dHdt}  
\end{figure}

There exists many different integration algorithms for the translation
and rotation variables. The choice of one algorithm over another
depends on the forces, the system size and the total simulation
time. In order to choose one integration algorithm for the
translational variables we investigated the conservation of energy in a
system of 50 particles colliding with the Hertz force in a rectangular
box. The implemented translational integration algorithms were:
Verlet~\cite{rapaport,allen}, Leap-Frog~\cite{rapaport,allen},
optimized Velocity Verlet ~\cite{omelyandecopt03,velverletopt} and a fifth-order
predictor-corrector method~\cite{rapaport}. The optimized velocity
Verlet method is written as
\begin{equation}
  \Vec R_1 = \Vec R(t) + \Vec V(t)\xi\dt,
\end{equation}
\begin{equation}
  \Vec V_1 = \Vec V(t) + \frac{1}{m}\Vec F[\Vec R_1]\dt / 2,
\end{equation}
\begin{equation}
  \Vec R_2 = \Vec R_1 + \Vec V_1(1-2\xi)\dt,
\end{equation}
\begin{equation}
  \Vec V(t+\dt) = \Vec V_1 + \frac{1}{m}\Vec F[\Vec R_2] \dt /2,
\end{equation}
\begin{equation}
  \Vec R(t+\dt) = \Vec R_2 + \Vec V(t+\dt)\xi\dt,
\end{equation}
where $\Vec R, \Vec V$ and $\Vec F$ represents the particle position,
velocity and force, respectively. The parameter \mbox{$\xi \simeq
  0.193183325037836$}\  minimizes the total truncation error of the
algorithm~\cite{omelyandecopt03}. For each time step, two computations of
the forces are needed, but even doubling the time step makes the
errors three times smaller than in the original Velocity Verlet
algorithm ($\xi = 0)$.  Figure~\ref{fig:dHdt} shows the variation of
the total mechanic energy $\langle \delta E^2 \rangle^{1/2} = (\langle
E^2 \rangle - \langle E \rangle^2 )^{1/2}$, averaged over the total
simulation time, as a function of the time step \dt. It is clear that
for large \dt the best algorithm is the optimized velocity Verlet
method. This is the algorithm we chose for the translational motion.

The simulation of spatial rotations is more complicated. It is well
known that the Euler angles can keep the information of a rigid body
on space, but there are some numerical instabilities on using
them~\cite{rapaport, allen}. For this reason, the Euler angles are
replaced by unitary quaternions, denoted here by $\Q$, and the
numerical problems are solved.  The main disadvantages on using
quaternions are the absence of high order integration algorithms and
the need to normalize the quaternion at each time step. These problems
are solved in a new formulation proposed by Omelyan~\cite{omelyan} of
the Leap-Frog method for quaternions that preserves the quaternion
unit norm, despite the size of the discrete time step.

In the original paper, the Omelyan algorithm is formulated in the $y$
convention of the Euler angles. We rewrote it on the $x$ convention,
\begin{equation}\label{equ:omelyan1}
  \begin{pmatrix}
    \dot\qcero\\
    \dot\quno\\
    \dot\qdos\\
    \dot\qtres\\
  \end{pmatrix}
  = \frac{1}{2}
  \begin{pmatrix}
    0 & -\omega_x^b & -\omega_y^b & -\omega_z^b\\
    \omega_x^b & 0 & \omega_z^b & -\omega_y^b \\
    \omega_y^b & -\omega_z^b & 0 & \omega_x^b \\
    \omega_z^b & \omega_y^b & -\omega_x^b & 0\\
  \end{pmatrix}
  \begin{pmatrix}
    \qcero \\
    \quno \\
    \qdos\\ 
    \qtres \\
  \end{pmatrix} \equiv Q(\vec\omega)\Q, 
\end{equation}
while the new angular velocities and quaternions are computed as
\begin{equation}\label{equ:omelyanomega}
  \vec\omega^b\left(t + \frac{\dt}{2}\right) = \vec\omega^b\left(t -
    \frac{\dt}{2}\right) + 
  \frac{\dt}{I}\vec\tau^b(t) + O(\dt^3),
\end{equation}
\begin{equation}\label{equ:omelyancuaternion}
  \Q(t+\dt) = \frac{I\left(1-\frac{\dt^2}{16}\omega^2\right) +
    \dt Q(\vec\omega)}{1 + \frac{\dt^2}{16}\omega^2}\Q(t) + O(\dt^3),
\end{equation}
where the superscript $^b$ represents the body fixed reference axes,
$I$ is the identity matrix on $R^4$, $Q(\vec\omega)$ is the matrix
defined on \eqref{equ:omelyan1} and the angular velocity  
$\vec\omega$ is computed in $t + \dt/2$.

\section{Simulation of the oedometric test}\label{SimOedometric}
Several element tests are performed on a soil in order to measure its
macroscopic parameters. An oedometric test \cite{evesque02} consists
on filling a metallic cylinder with a sample of the soil and measuring
the initial void ratio $e_0$, that is the ratio between the volume
occupied by voids over the total volume of the sample. All cylinder
walls but the top one are fixed. Then, an initial vertical pressure
$p_0$ is applied on the top wall and, after some time (typically 10
minutes), the sample relaxes to a new void ratio. Then, the pressure
is doubled, the sample relaxes and a new void ratio is measured.  The
procedure repeats until reaching some maximal pressure before the
particles crush.

\begin{figure}[ht]
  \centering
  \includegraphics[scale=0.35]{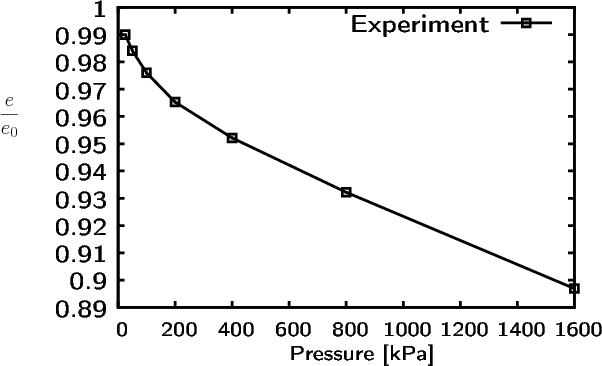}
  \vbox{\strut{\phantom{pb}}}
  \includegraphics[scale=0.35]{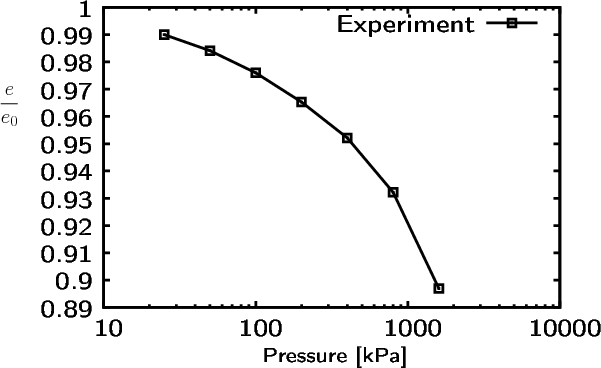}
  \caption{Experimental oedometric test results on a dry Guamo sand in
    (top) linear and (bottom) log-scale on the applied vertical
    stress.}\label{fig:expRes}
\end{figure}

The curve so obtained (figure~\ref{fig:expRes}) can be fit by the
Bauer empirical equation~\cite{hypo07} 
\begin{equation}
  \frac{e}{e_0} = \exp \left[ -\left( \frac{p}{\hs} \right) ^\n
  \right]\quad 
\end{equation}
by using, for instance, the Levenberg-Marquardt algorithm \cite{nr}.
This gives the two hypoplastic parameters \hs and \n.
Figure~\ref{fig:expRes} shows the oedometric curve for a Guamo
sand with grains between 0.85mm and 1.15mm diameter; we obtained
$\hs[\text{MPa}]=19.1(1.1)$ and $\n = 0.57(1)$.

The same experimental procedure has been implemented in the
simulations. For each set of microscopic parameters, the grains are
randomly distributed inside the cylindrical container; they fall down
by gravity and collide each other and with the walls until rest.
Then, the top wall falls down with the initial pressure $p_0$ and the
oedometric test starts, following the same procedure than in the
experimental test.  We try the simulation to be as similar to
the experiment as possible: The dimensions of the container, the
polydispersity of the sample and others parameters have almost the
same experimental values.  The only difference is in the number of
grains: around 5000 for the experiment and 280 for the simulation
(because of hardware limitations).  At a first glance it appears as a
very drastic reduction, but the results were very closed to the
experimental data, and it gives us some confidence on the procedure. 
Moreover, current simulations with more than 1000 particles shows
similar behaviour to the present ones. The typical computational time
for the simulation of a complete oedometric test was about 52 hours on
a Pentium IV 3.0 GH machine with 4GB of RAM. The compiler used was
gnu/g++. That was done for $\delta t \simeq 10^{-7}$
seconds\footnote{Currently, by using the same theoretical model and
  some special compiler flags, we are able to simulate the oedometric
  test in about three days with 1200 particles.}.

\begin{figure}[ht]
  \centering
  \includegraphics[scale=0.36]{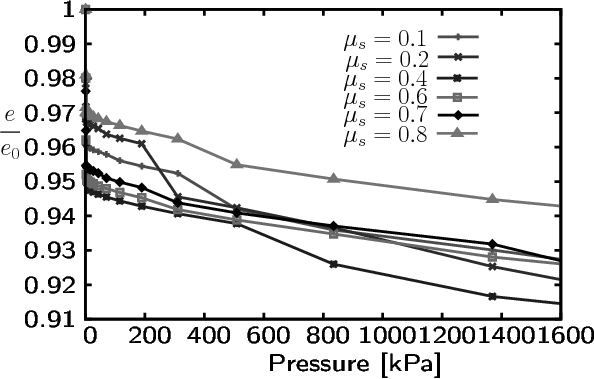}\qquad
  \includegraphics[scale=0.36]{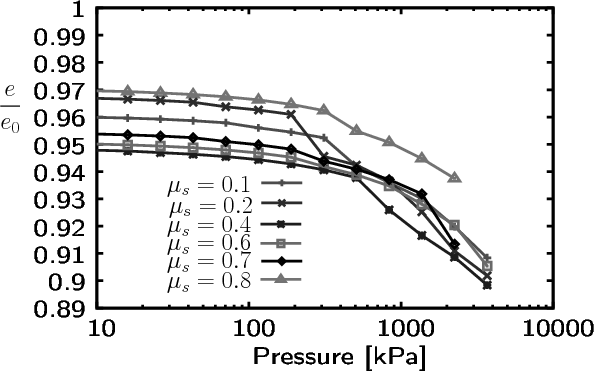}
\caption{Void ratio versus vertical stress for various values of
  the microscopic  static
  friction coefficient $\mu_s$.}\label{fig:simSampleCurve} 
\end{figure}

\begin{table}[!ht]
  \centering
  \begin{tabular}{ccc}\hline\hline
    $\boldsymbol{\V \text{[GPa]}}$ & 
    $\boldsymbol{\n}$ & $\boldsymbol{\hs \text{[MPa]}}$\\\hline\hline
    0.0112 & 11(3) $\times 10^{-1}$& 1(2) $ \times 10^{-1}$\\
    0.1125 & 10(3) $\times 10^{-1}$& 6(6) $\times 10^{-1}$\\
    1.1250 & 51(3) $\times 10^{-2}$  & 34(5) \\
    18.000 & 24(2) $\times 10^{-2}$  & 5(2) $\times 10^{4}$ \\\hline
  \end{tabular}\\\vspace*{1ex}
  \begin{tabular}{ccc}\hline\hline
    $\boldsymbol{\uk}$ & 
    $\boldsymbol{\n}$ & $\boldsymbol{\hs \text{[MPa]}}$\\\hline\hline
    0.1 & 428(7)  $\times 10^{-3}$  & 176(7) $\times 10^{1}$\\
    0.2 & 302(3)  $\times 10^{-3}$ & 159(6) $\times 10^{2}$\\
    0.4 & 26(2) $\times 10^{-2}$ & 4(1) $\times 10^{4}$\\
    0.6 & 40(7) $\times 10^{-2}$ & 13(7) $\times 10^{2}$\\\hline
  \end{tabular}
  \caption{Macroscopic parameters $\hs$ and $\n$ for different values of
    the microscopic parameter  $V$ (top) and $\uk$ (bottom).}\label{tbl:macroMicrovuk}  
\end{table}  

Two typical curves for the simulation of the oedometric compression
are  shown in Figure~\ref{fig:simSampleCurve}. Similar curves are
obtained for the other parameters by  following the same procedure as before 
(see Tables~\ref{tbl:macroMicrovuk} and \ref{tbl:macroMicrogammanus}
for details).

\begin{table}[!ht] 
  \centering
  \begin{tabular}{ccc}\hline\hline
    $\boldsymbol{\gamman [\text{s}^{-1}\text{m}^{-1/2}]}$ & 
    $\boldsymbol{\n}$ & $\boldsymbol{\hs \text{[MPa]}}$\\\hline\hline
    5.0e1 & 287(1) $\times 10^{-3}$&  250(4) $\times 10^{2}$\\
    5.0e2 & 29(4) $\times 10^{-2}$& 16(8) $\times 10^{3}$\\
    5.0e3 & 32(2) $\times 10^{-2}$& 8(1) $\times 10^{3}$\\
    5.0e4 & 27(4) $\times 10^{-2}$& 2(1) $\times 10^{4}$\\
    5.0e6 & 26(3) $\times 10^{-2}$& 4(2) $\times 10^{4}$\\
    5.0e7 & 20(4) $\times 10^{-2}$& 7(6) $\times 10^{4}$\\\hline
  \end{tabular}\vspace*{1ex}
  \begin{tabular}{ccc}\hline\hline
    $\boldsymbol{\us}$ & 
    $\boldsymbol{\n}$ & $\boldsymbol{\hs \text{[MPa]}}$\\\hline\hline
    0.1 & 26(2) $\times 10^{-2}$& 32(7) $\times 10^{3}$\\
    0.2 & 30(2) $\times 10^{-2}$& 7(2)$\times 10^{3}$ \\
    0.4 & 24(1) $\times 10^{-2}$& 4(1) $\times 10^{4}$\\
    0.6 & 20(2) $\times 10^{-2}$& 5(3)$\times 10^{5}$ \\
    0.7 & 19(3) $\times 10^{-2}$& 10(8) $\times 10^{5}$\\
    0.8 & 23(1) $\times 10^{-2}$& 5(2) $\times 10^{5}$\\\hline
  \end{tabular}
  \caption{Macroscopic parameters $\hs$ and $\n$ for different values of
    the microscopic parameter $\gamman$ (top) and  $\us$
    (bottom).}\label{tbl:macroMicrogammanus}   
\end{table}

\begin{figure}[!ht]
  \centering
  \includegraphics[scale=0.40]{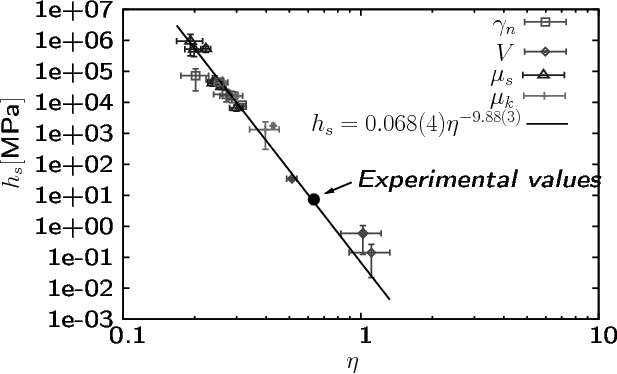}
  \caption{\hs as a function of \n for different values of the
    microscopic parameters}\label{fig:hsVsn}
\end{figure}

We noted that, in all cases, the two hypoplastic parameters behave in
a non-independent way. When one parameter increases, the other one
decreases. This suggest a hidden relationship between them.  In order
to investigate this relation, we plot \hs as a function of \n for all
sets of microscopic parameters, as shown in figure~\ref{fig:hsVsn}. It
reveals and unsuspected power law relationship between the two
hypoplastic parameters: $\hs$$=$$0.068(4)\n^{-9.88(3)}$. Moreover, the
experimental values lay on the line. This kind of relation could
suggest a possible reduction in the number of macroscopic parameters
of the hypoplastic theory, but the experimental values for different
sands with different granulometries do not lay on a single power law.
Let us point out that we have varied all microscopic parameters ($V$,
$\gamma_n$, $\mu_k$ and $\mu_s$), but the granulometry. It is well
known~\cite{nccfromexp} that \n strongly depends on the granulometry,
but once this variable is fixed, a power law could appear. This result
should be validated by future experiments and simulations.

\section{Conclusions}\label{Conclusions}

Hereby we have simulated the oedometric compression of a bad-graded
Guamo sand in order to estimate the hypoplastic parameters \hs and \n,
and we have compared with experimental results.  In particular, we
have investigated the changes on these two macroscopic parameters by
varying four microscopic parameters: the stiffness $V$, the normal
damping constant $\gamma_n$, the static friction coefficient  $\mu_s$
and dynamic friction coefficient $\mu_k$.  By doing so, we found
an unexpected power-law relationship between \hs and \n: for our case,
$\hs \propto \n^{-9.88(3)}$. Moreover, the experimental values for the
Guamo sand with the same granulometry (random-distributed diameters
between 0.85 and 1.15mm) lies on the same curve. This suggests us that
these two parameters may be replaced by other two, more related with
the microscopic world: one reflecting the granulometry and another one
reflecting the strength of the microscopic interactions. This power
law, of course, must be confirmed by a broader set of experiments, but
the possibility is very promising. The experimental confirmation of
these kind of relationships and the possible definition of these two
new parameters are subjects of present research.

The simulation performed is 3D and the software developed by us
includes some of the state-of-the-art algorithms for spatial
translation and 
rotations. Moreover, the microscopic force model includes
rolling among the more traditional elastic and damping normal
interactions and sliding and static frictional forces.  The rolling
process allows global dissipation and long-range reorientations without
requiring a big spatial reconfiguration of the sample, that is without
changing the macroscopic void ratio, and its global effect on the
macroscopic parameters is very interesting to explore.  For example,
it would be possible to perform simulations with and without the
rolling force in order to get insight into its actual role on the soil
properties.  This will be a topic of future work.

Micro-mechanical simulations are powerful tools for the investigation
on the microscopic origin of macroscopic behavior of
soils. Furthermore, these computer experiments have shown to be able
to approximate the experimental parameters for the soil. We hope this
work will raise new questions regarding the microscopic origin of the
constitutive parameters and gives starting points for a redefinition
of the macroscopic parameters, more related with the microscopic
world.




\bibliography{biblio}
\end{document}